%% file: main.tex
\documentclass{article}
\usepackage[utf8]{inputenc}
\usepackage{comment}
\usepackage{amsmath, amssymb}
\usepackage{pdfpages}
\usepackage{lmodern}
\usepackage[hidelinks]{hyperref}
\usepackage[a4paper, margin=2.5cm]{geometry}
\usepackage[linesnumbered, spanish, ruled]{algorithm2e}
\usepackage{siunitx}
\def\sym#1{\ifmmode^{#1}\else\(^{#1}\)\fi}
\usepackage{tikz}
\usetikzlibrary{calc}
\usepackage{graphicx}
\usepackage{float}
\usepackage{bm}
\usepackage{longtable} 
\usepackage{pdflscape} 
\usepackage[skip=7pt plus1pt, indent=20pt]{parskip}

\newrobustcmd{\orcid}[1]{%
  \href{https://orcid.org/#1}{\includegraphics[height=1.7ex]{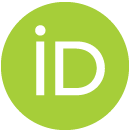}%
  \,https://orcid.org/#1}}

\usepackage{setspace}
\usepackage{tabularx}
\usepackage{makecell}
\usepackage{mathrsfs}
\usepackage{caption}
\usepackage{subcaption}
\usepackage[utf8]{inputenc}
\usepackage{rotating}
\newcolumntype{L}[1]{>{\center\arraybackslash} m{#1} }
\usepackage[nolist]{acronym}
\input{bibliography.tex}

\begin{document}
\pagenumbering{arabic}
    \begin{acronym}
        \acro{ANSES}{Administración Nacional de la Seguridad Social}
        \acro{CUS}{Cobertura Universal en Salud}
        \acro{EED}{Experimento de Elección Discreta}
        \acro{ENGHo}{Encuesta Nacional de Gastos de los Hogares}
        \acro{FUA}{Función de Utilidad Aleatoria}
        \acro{GCS}{gasto catastrófico en salud}
        \acro{GBS}{gasto de bolsillo en salud}
        \acro{INDEC}{Instituto Nacional de Estadística y Censos}
        \acro{INSSJyP}{Instituto Nacional de Servicios Sociales para Jubilados y Pensionados}
        \acro{LAC}{Latinoamérica y el Caribe}
        \acro{MLM}{Modelo Logístico Multinomial}
        \acro{MLMMX}{Modelo Logístico Multinomial Mixto}
        \acro{LV}{logaritmo de la función de verosimilitud}
        \acro{LR}{test de cociente de verosimilitud}
        \acro{MEDICARE}{Medical Care}
        \acro{MG}{médico generalista}
        \acro{OMS}{Organización Mundial de la Salud}
        \acro{ONU}{Organización de las Naciones Unidas}
        \acro{OR}{odd ratio}
        \acro{PAMI}{Programa de Atención Médica Integral}
        \acro{RDD}{análisis de regresión discontinua}
        \acro{RKD}{análisis de discontinuidad en la pendiente}
        \acro{RPLP}{análisis de regresión polinómica local ponderada}
        \acro{SIPA}{Sistema Integrado Previsional Argentino}
        \acro{SSA}{sistema de salud argentino}
    \end{acronym}

\doublespacing
\begin{titlepage}
\pagenumbering{gobble}
\begin{center}
\Huge{Evaluación del efecto del PAMI en la cobertura en salud de los adultos mayores en Argentina}\\[1cm]
\Large{Juan Marcelo Virdis\textsuperscript{1*}, Fernando Delbianco\textsuperscript{2, 3} y María Eugenia Elorza\textsuperscript{1, 2}}\\[.5cm]
\end{center}

\begin{center}\small{
\textsuperscript{1}Instituto de Investigaciones Económicas y Sociales del Sur, Concejo Nacional de Investigaciones Científicas y Técnicas (CONICET) - Universidad Nacional del Sur (UNS), San Andres 800, Bahía Blanca, Argentina. \\
\textsuperscript{2}Departamento de Economía, UNS, San Andrés 800, Bahía Blanca, Argentina.\\
\textsuperscript{3}Instituto de Matemática de Bahía Blanca, CONICET-UNS,  Av. Alem 1253, Bahía Blanca, Argentina
}\\
\end{center}

\vspace{3mm}
\small{
\noindent    Nota del autor\\
\indent     Juan Marcelo Virdis     \orcid{0000-0001-7118-9259}\\
\indent     Fernando Delbianco      \orcid{0000-0002-1560-2587} \\
\indent     María Eugenia Elorza    \orcid{0000-0003-1562-1363}\\
\\
\textsuperscript{*} Autor para correspondencia. Información de contacto: \\
\noindent Instituto de Investigaciones Económicas y Sociales del Sur (CONICET-UNS), Departamento de Economía, Universidad Nacional del Sur. \\ 
San Andrés 800, Bahía Blanca, Buenos Aires, Argentina (CP: 8000) \\
E-mail: jmvirdis@iiess-conicet.gob.ar \\
Teléfono: +54 291 4595138
}
\pagebreak
\end{titlepage}

\pagenumbering{arabic}

\section{Introducción}\label{sec:rdd.introduccion}
    
    El \ac{PAMI} es una institución cuyo estudio es relevante para evaluar el desempeño del \ac{SSA}.
    Entre otros aspectos, se puede destacar que otorga un seguro de salud a cinco millones de personas aproximadamente, que su población objetivo es la que mayor gasto esperado tiene, los adultos mayores, y que la transición demográfica que se atraviesa Argentina causará que, de mantenerse la legislación actual, el numero de beneficiarios se incremente, tanto en términos absolutos como relativos dentro del \ac{SSA}. 
    
    Uno de los aspectos relevantes a evaluar es el impacto que tiene el \ac{PAMI} en la cobertura en salud de sus beneficiarios, entendida como el acceso a la salud y la protección financiera contra el \ac{GBS}. 
    Este tipo de investigaciones puede realizarse a través de distintas clases de regresiones, utilizando como variable dependiente indicadores de protección financiera o el \ac{GBS} de los hogares.
    La dimensión del acceso podría evaluarse utilizando variables categóricas que indiquen si los hogares o las personas han utilizado servicios de salud preventivos y de tratamiento (\ac{OMS} \& Grupo del Banco Mundial, \citeyear{Organizacion_Mundial_de_la_Salud2014-oq}).
    Sin embargo, tanto para variables de protección financiera como de acceso, el uso de datos obtenidos en encuestas puede resultar en estimaciones sesgadas, a pesar de que los datos correspondan a una muestra probabilística representativa de la población.
    Esto se debe a la dificultad para observar el estado de salud de las unidades observacionales, aun en encuestas que incluyen preguntas relacionadas.
    Distintos autores han encontrado diferencias entre el autoreporte y la evaluación objetiva de hipertensión, diabetes, asma y enfermedades cardíacas \parencite{Mulyanto2019-al, Okura2004-ye, Tenkorang2015-lz}. Además, los participantes de encuestas pueden ser inconsistentes cuando clasifican su salud entre distintas categorías. Se ha encontrado evidencia de que las respuestas pueden diferir según la naturaleza del relevamiento (verbal o escrito) o la secuencia en que estén ordenadas las preguntas \parencite{Crossley2002-hu}.
    El control de variables relacionadas a la salud es muy importante por su incidencia en la demanda de seguros de salud y en la demanda de bienes y servicios relacionados con la salud.
    De otra manera, podría producirse un problema de endogeneidad en la estimación \parencite{Levy2008-zt, Trujillo2003-va, Vera-Hernandez1999-ym, Waters1999-od}.
    

    
    Una solución al problema de endogeneidad es el uso de técnicas cuasiexperimentales, las cuales consisten en conformar a partir de datos de encuestas un grupo de estudio y un grupo control, seleccionando unidades observacionales iguales en todas sus características, a excepción de una variable tratamiento y una variable objetivo \parencite{Chapin1938-pm, Thistlethwaite1960-ax}.
    De esta forma, es posible evaluar la existencia de una relación causal entre ellas.
    Cualquier otra variable que sea diferente entre el grupo de estudio y el grupo control debilitará los resultados, pues las variables no controladas pueden haber incidido en la variable objetivo. 
    En estudios basados en información proveniente de encuestas, las variables que es posible controlar estarán limitadas por los datos que es posible evaluar a partir de un cuestionario.
    Por esta razón, siempre surgirá el interrogante sobre la existencia de sesgos en las estimaciones producidos por variables no observadas o de difícil medición. 
    
    Una metodología cuasiexperimental que puede atenuar el sesgo por variables no controladas es el \ac{RDD} \parencites{Thistlethwaite1960-ax}    \footnote{Otras metodologías cuasiexperimentales utilizadas en la evaluación de efectos causales pueden consultarse en \textcite{Cunningham2021-uf}.}.
    El \ac{RDD} consiste en observar el cambio en una variable objetivo causada por un tratamiento, el cual es asignado a partir de un valor determinado o punto de corte de una variable continua observable.
    De esta forma, las unidades observacionales que superan el punto de corte habrán recibido el tratamiento, a diferencia de las que no lo han hecho.
    La ventaja del \ac{RDD} reside en que las observaciones que se encuentran lo suficientemente cerca del punto de corte serían similares, lo que permite reducir la cantidad de variables control necesarias.
    
    Esta metodología ha sido aplicada para evaluar el efecto de seguros en la cobertura en salud en distintas partes del mundo.
    \textcite{Bernal2017-if} encontraron que un seguro gratuito existente en Perú ha causado mayor acceso a servcios de salud, pero también genero un mayor \ac{GBS}.
    Por otra parte, \textcite{Palmer2015-yp} encontraron que un seguro para niños menores a los 6 años en Vietnam genera mayor utilización de servicios ambulatorios y de medicina interna, sin encontrar modificaciones en el \ac{GBS}.
    En el caso de \textcite{Card2008-vv} y \textcite{Card2009-vx} evaluaron el impacto del programa \ac{MEDICARE} de Estados Unidos en el acceso a prestaciones médicas y en los resultados de salud de los adultos mayores.
    En estos estudios, se evaluaron cambios en las variables objetivo cuando las personas llegaron a los 65 años de edad, momento en el cual es posible para los estadounidenses que han trabajado 40 trimestres acceder de forma gratuita a \ac{MEDICARE}.
    Los autores encontraron aumentos en el porcentaje de la población con seguro, las consultas médicas y en otras prestaciones.
    Además, se encontró una reducción del 20 \% en la tasa de mortalidad de pacientes severos a partir de los 65 años.
    
    Al igual que en los trabajos descriptos, la normativa legal existente para acceder al PAMI permite evaluar efectos causales de este seguro sobre la cobertura en salud a través de un \ac{RDD} utilizando como grupo de estudio a las personas que recientemente alcanzaron la edad jubilatoria y como grupo control a quienes están cerca de hacerlo.
    Es probable que estos grupos sean lo suficientemente similares en muchos factores que se encuentran invariables al momento de cruzar el umbral de la edad jubilatoria, entre los que se incluye el estado de salud.

    Este capítulo tiene por objetivo general evaluar el efecto del \ac{PAMI} en la cobertura en salud. 
    Como objetivos específicos se propone estimar el impacto del \ac{PAMI} en la utilización de servicios médicos, en el \ac{GBS}, en indicadores de protección financiera contra \ac{GBS} y en la demanda de seguros voluntarios.
    Los resultados obtenidos a partir de esta evaluación permitirán diagnosticar la eficacia del \ac{PAMI} como financiador de servicios de salud para los adultos mayores y orientar el diseño de políticas por parte del organismo en relación a su oferta prestacional.
    Los resultados podrían implicar la necesidad de ampliar la oferta de servicios o mejorar los mecanismos de acceso que deben utilizar los beneficiarios. 
    En la segunda sección se describen los datos y se expone la metodología utilizada. En la tercer sección se presentan los resultados. En la cuarta sección, se presenta una discusión de los resultados.

    \section{Datos y metodología}\label{sec:rdd.datos.met}
    
    Las estimaciones que se presentan en este capítulo fueron realizadas con datos de la \ac{ENGHo} relevados entre los años 2017 y 2018 (\ac{INDEC}, \citeyear{Instituto_Nacional_de_Estadistica_y_Censos2020-iq}).
    La \ac{ENGHo} es un relevamiento que permite conocer las estructuras de gasto de los hogares y caracterizar a la población a través de variables socioeconómicas (\ac{INDEC}, \citeyear{Instituto_Nacional_de_Estadistica_y_Censos2020-cx}). 
    La encuesta fue realizada entre los años 2017 y 2018 a una muestra diseñada en un procedimiento de tres etapas realizado sobre hogares situados en localidades de 2000 habitantes o más%
    \footnote{Para un mayor detalle del diseño muestral ver \ac{INDEC} (\citeyear{Instituto_Nacional_de_Estadistica_y_Censos2020-vh}).}. 
    La muestra estuvo compuesta por 44.922 viviendas de las cuales se obtuvieron 21.547 respuestas, lo que abarca a 68.725 habitantes. 
    Cada uno de los hogares fue asociado a factores de expansión que permiten ajustar mediciones estadísticas por no respuesta, vivienda no elegible y calibración por \textit{benchmarks} o totales poblacionales conocidos (\ac{INDEC}, \citeyear{Instituto_Nacional_de_Estadistica_y_Censos2020-vh}). 
    Las bases de datos fueron obtenidas en el sitio web oficial del \ac{INDEC} (\citeyear{Instituto_Nacional_de_Estadistica_y_Censos2020-iq}).
    La unidad observacional es el hogares dado que en las bases de datos de la \ac{ENGHo} 
    es imposible conocer cuál de los integrantes accedió al sistema de salud o realizó \ac{GBS}.
    Esto conlleva la tarea de asociar una edad, género y/o situación laboral a hogares compuestos por integrantes con distintos grados de heterogeneidad.
    Para abordar este obstáculo, se estableció el supuesto de que el hogar es caracterizado por las variables relevadas para su jefe de hogar\footnote{
    En la \ac{ENGHo} el jefe de hogar  es la persona considerada como tal por los demás integrantes del hogar. 
    En cada hogar hay un solo jefe, por lo tanto, hay tantos jefes como hogares (\ac{INDEC}, \citeyear{Instituto_Nacional_de_Estadistica_y_Censos2022-tt}).}.
    Este supuesto se fundamenta en la legislación que reglamenta a las Obras Sociales y al \ac{PAMI}, que establece que cada beneficiario titular del seguro de salud cuenta con la posibilidad de extenderlo a su grupo familiar (Ley n° 19.032, art. 2, \citeyear{Ley_n_19032_Instituto_Nacional_de_Servicios_Sociales_parar_Jubilados_y_Pensionados_Art_21971-mz}; Ley n° 23.660, art. 9, \citeyear{Ley_n_23660_Obras_Sociales_Art_91988-el}).
    Por este motivo, el hecho que el jefe de hogar cumpla la edad necesaria para poder acceder a los beneficios del \ac{PAMI}, podría tener un efecto en el seguro de salud de todos los integrantes del hogar y, por ende, en la cobertura en salud.

    
    A continuación se describe el \ac{RDD} en su forma no paramétrica siguiendo la notación de \textcite{Hahn2001-ov}.
    Se define a $x_{i}$ como una variable binaria sobre la cual deseamos evaluar el efecto en una variable objetivo $y_{i}$.
    La variable $x_{i}$ es llamada tratamiento e indica si la $i-esima$ observación lo ha recibido.
    El valor de la variable objetivo puede ser expresado como
    \begin{equation*}
        y_{i} = \alpha_{i}+x_{i}\beta_{i}
    \end{equation*}
    
    donde $\alpha_{i}\equiv y_{0i}$ y $\beta_{i}\equiv y_{1i}-y_{0i}$.
    La identificación de las unidades observacionales que han recibido el tratamiento es realizada a partir de una variable continua $z_{i}$.
    En un punto de corte $z_{i}=z_{0}$ se produce un hecho exógeno que indica que alguna proporción de las observaciones ha recibido el tratamiento $x_{i}$.
    En los \ac{RDD} realizados en este capítulo se utilizaron las variables que se presentan en la Tabla \ref{tab:rdd.vars}. La variable $z_{i}$ se define como la edad del jefe de hogar menos la edad jubilatoria del régimen general, la cual es 65 años para varones y 60 años para mujeres (Ley n° 24.241 art. 19, \citeyear{Ley_n_24241_Sistema_Integrado_de_Jubilaciones_y_Pensiones_Art_191993-xt}). 
    A la edad jubilatoria de ambos sexos se le sumó un año dado que existe una demora de entre entre tres y seis meses para finalizar el trámite administrativo de la jubilación, momento a partir del cual es posible iniciar la gestión para obtener \ac{PAMI} (\cite{Ambito_Financiero2022-bp}; \ac{INSSJyP}, \citeyear{Instituto_Nacional_de_Servicios_Sociales_para_Jubilados_y_Pensionados2022-mw}).
    De esta forma, la variable continua observada queda definida como:
    \begin{equation*}
    z_{i} =
    \begin{cases}
        edad_{i}-61 \: \text{ si } sexo_{i}=1 \\
        edad_{i}-66 \: \text{ si } sexo_{i}=0
    \end{cases}
    \end{equation*}
    
    \begin{table}[htp!]
        \centering
        \caption{Variables utilizadas en las estimaciones de la sección \ref{sec:rdd.resultados}}
        \input{tablas/rdd.vars}
        \label{tab:rdd.vars}
    \end{table}
    
    En la literatura existen dos tipos de \ac{RDD}: los \textit{sharp} y los \textit{fuzzy}. 
    En el diseño \textit{sharp}, todas las observaciones para las cuales $z_{i}\geq z_{0}$ reciben el tratamiento. Es decir, $x_{i}=f(z_{i})$, donde $z_{i}$ es continua y $f(z_{i})$ es discontinua en un valor conocido $z_{0}$.
    En el diseño \textit{fuzzy}, $x_{i}$ es una variable aleatoria con una función de probabilidad condicional $f(z)\equiv E[x_{i}|z_{i}=z] = Pr[x_{i}=1|z_{i}=z]$, la cual presenta una discontinuidad en $z_{0}$.
    La diferencia entre estos diseños radica en que en el \textit{sharp} la variable $z_{i}$ indica de forma determinística las unidades observacionales que han recibido un tratamiento mientas que en el \textit{fuzzy} esta variable indica un cambio en la probabilidad de recibir el tratamiento. 
    Ambos diseños comparten los siguientes supuestos:
    \begin{enumerate}
        \item [i:] Existen límites $x^{+}\equiv \lim_{z\rightarrow z_{0}^{+}} E[x_{i}|z_{i}=z]$ y $x^{-}\equiv \lim_{z\rightarrow z_{0}^{-}} E[x_{i}|z_{i}=z]$
        \item[ii:] $x^{+}\neq x^{-}$
    \end{enumerate}
    Estos supuestos implican que es posible acercarse infinitesimalmente al punto de corte, tanto desde valores mayores como de valores menores, y que existe una discontinuidad en el valor esperado del tratamiento en $z_{0}$.
    De forma no-paramétrica puede definirse el efecto del tratamiento en un diseño \textit{fuzzy} como
    \begin{equation}
        \hat{\tau}_{RD} = E[\beta_{i}] = \frac{y^{+}-y^{-}}{x^{+}-x^{-}}
        \label{eq:rdd.fuzz.np}
    \end{equation}
    siendo el diseño \textit{sharp} un caso particular en el que $x^{+}=1$, $x^{-}=0$ y $\hat{\tau}=y^{+}-y^{-}$.
    El parámetro $\hat{\tau}$ estimado es el salto que se verificaría en la variable objetivo $y$ si el 100 \% de las unidades observacionales hubiera recibido el tratamiento $x$.

    El efecto estimado del tratamiento a partir de la ecuación \ref{eq:rdd.fuzz.np} será correcto siempre que no haya otra diferencia entre los grupos tratamiento y control que afecte a la variable objetivo $y$, lo cual puede resultar un supuesto fuerte para las características del efecto que se intenta medir en este capítulo.
    Esto sucede porque al momento de jubilarse los individuos pueden tener cambios importantes en otras variables relevantes para la adquisición de un seguro de salud y el consumo de bienes y servicios de salud, como el nivel de ingreso o la situación laboral, variable importante en el costo de oportunidad del tiempo.
    A su vez, estas variables pueden ser distintas entre hombres y mujeres producto de las diferencias de género existentes en el mercado laboral.
    Este problema puede ser abordado a través de la inclusión de covariables en el \ac{RDD} lo cual, según investigaciones previas, puede mejorar la precisión de los resultados \parencite{Calonico2019-bs, Frolich2007-xp}.
    Por esta razón, se incluyeron como covariables $sexo_{i}$, $lgasto_{i}$ e $inac_{i}$ (ver descripción en Tabla \ref{tab:rdd.vars}).
    

    Además de la discontinuidad en la variable objetivo, se evaluó la existencia y magnitud de un cambio de pendiente alrededor del punto de corte.
    De esta manera, es posible observar si se modifica la dinámica entre las variables objetivo y la edad del jefe de hogar.
    Este tipo de estimación se denomina \ac{RKD} y, a diferencia del \ac{RDD} en el cual se evalúan diferencias en los niveles de la variable objetivo alrededor de $z_{0}$, en el \ac{RKD} se estiman diferencias en la primer derivada de la función de regresión.
    
    Para la estimación de la ecuación \ref{eq:rdd.fuzz.np} se utilizó la metodología desarrollada por \textcite{Calonico2014-tl}.
    Los autores proponen estimar un \ac{RPLP} de orden 1 o 2 para aproximar los valores esperados por debajo y por encima del límite $z_{0}$ en una variable objetivo $y$.
    La \ac{RPLP} es una extensión de las estimaciones paramétricas que permite ajustar una curva suavizada para $y=f(z)$ \parencite{Cleveland1996-sn}.
    Se define a la función de regresión $m(\cdot)$ como
    \begin{equation*}
    \hat{m}_{h}(x) =
    \begin{cases}
        \hat{\alpha}_{-}(z) \text{     si } z<z_{0} \\
        \hat{\alpha}_{+}(z) \text{     si } z\geq z_{0}
    \end{cases}
    \end{equation*}
    en la cual los parámetros estimados $\hat{\alpha}_{-}$ y $\hat{\alpha}_{+}$ surgen de resolver por mínimos cuadrados las siguientes expresiones
    \begin{equation*}
    \hat{\mu}(z)=
    \begin{cases}
        \left(\hat{\alpha}_{-}(z), \hat{\beta}_{-}(z)\right)= \arg\,\min\limits_{\alpha,\beta}\sum_{i=1}^{N}\textbf{1}_{z_{i}<z}\left(y_{i}-\alpha-\sum_{1}^{p}\beta_{q}(z_{i}-z)^{p}\right)^{2}w(z_{i}) \\
        \left(\hat{\alpha}_{+}(z), \hat{\beta}_{+}(z)\right)= \arg\,\min\limits_{\alpha,\beta}\sum_{i=1}^{N}\textbf{1}_{z_{i}>z}\left(y_{i}-\alpha-\sum_{1}^{p}\beta_{q}(z_{i}-z)^{p}\right)^{2}w(z_{i})
    \end{cases}
    \end{equation*}
    donde $q$ es el orden de la regresión. El efecto estimado es igual a
        \begin{equation*}
  \hat{\tau}_{RD}=\hat{\mu}_{+}-\hat{\mu}_{-} 
    \end{equation*}
    donde
        \begin{equation*}
    \hat{\mu}_{-} = \lim_{z\rightarrow z_{0}^{-}} \hat{m}_{h}(z)=\hat{\alpha}_{-}(z_{0})    
    \end{equation*}
        \begin{equation*}
    \hat{\mu}_{+} = \lim_{x\rightarrow z_{0}^{+}} \hat{m}_{h}(z)=\hat{\alpha}_{+}(z_{0})    
    \end{equation*}
    El intervalo de confianza de $\hat{\tau}_{RD}$ es igual a
        \begin{equation*}
    CI_{1-\alpha,n}^{rbc}=\left[\left\{\hat{\tau}_{RD}(h_{n})-\hat{b}_{n}\right\}\pm\Phi^{-1}_{1-\frac{\alpha}{2}}\sqrt{\hat{v}^{bc}_{n}}\right]
    \end{equation*}
    donde el superíndice $rbc$ denota que es un intervalo robusto con correcciones de sesgo $\hat{b}_{n}$ en la estimación de $\hat{\tau}(h_{n})$ y en la estimación de la varianza.
    Las correcciones de los sesgos se describen en \textcite{Calonico2014-bl}.
    
    La estimación de una \ac{RPLP} requiere determinar: i) el orden de la \ac{RPLP}, ii) el ancho de banda $h$, es decir, las observaciones 
    $(y_{i}, z_{i}):z_{i}\in [z_{0} - h ; z_{0} + h]$ 
    que se utilizarán en las \ac{RPLP} y iii) la función de ponderación $w(z_{i})$.
    En relación al orden, no es recomendable en \ac{RDD} realizar regresiones polinómicas de orden mayor a 2.
    Esto se debe a que la estimación de la diferencia en los valores esperados a la izquierda y a la derecha del punto de corte es sensible al orden del polinomio, y la utilización de polinomios de orden alto puede incrementar la probabilidad de error de tipo 1, es decir, encontrar significatividad estadística en una discontinuidad cuando esta no existe \parencite{Gelman2019-xa}.
    En relación a $w(z_{i})$, las estimaciones no-paramétricas pueden presentar estimaciones sesgadas cuando el soporte de la curva verdadera es acotado, el cual puede ser reducido si se utiliza una función de ponderación triangular \parencite{Cheng1997-jn}:
    \begin{equation}\label{eq:w.k}
        w_{k}=1-\frac{|x_{k}-x_{i}|}{h}
    \end{equation}

    

    Por último, los \ac{RDD} utilizando \ac{RPLP} son sensibles al $h$ utilizado. 
    En este sentido \textcite{Calonico2019-yy} sugieren estimarlo como:
    \begin{equation}
        h=\left[\frac{(1+2q) \hat{v}^{bc}_{n}}{2(1+p-q)\hat{b}_{n}}\right]^{1/(2p+3)}n^{-1/(2p+3)}
    \end{equation}
    donde $p$ es el orden de la \ac{RPLP}, $q$ es el orden de la derivada ($0$ en el caso de \ac{RDD} y $1$ en \ac{RKD}), $\hat{v}^{bc}_{n}$ la aproximación de la varianza y $\hat{b}_{n}$ la aproximación del sesgo en el estimador de discontinuidad.
    Con el fin de conocer la sensibilidad de los resultados a la especificación elegida, se realizaron estimaciones de \ac{RDD} y \ac{RKD} utilizando polinomios de primer y segundo orden, con y sin covariables.
    Dado que la edad jubilatoria no determina inequivocamente que las personas se han convertido en beneficiarias del \ac{PAMI}, todas las estimaciones realizadas son del tipo \textit{fuzzy}.
    Las estimaciones fueron realizadas con el paquete para RStudio \textit{Robust Data-Driven Statistical Inference in Regression-Discontinuity Designs} \parencite{Calonico2022-po}.
    

\section{Resultados}\label{sec:rdd.resultados}

    En la Tabla \ref{tab:rdd.est.desc} se presentan las estadísticas descriptivas de las variables utilizadas en este capítulo.
    Se observa que un 73 \% de los hogares tienen un jefe de hogar titular de algún tipo de seguro, mientras que el 27 \% no tiene seguro de salud.
    Además, un 20 \% de los jefes de hogar es beneficiario del \ac{PAMI}, mientras que
    los jefes de hogar con seguro voluntario o más de un seguro representan el 4,8 y 3,2 \% respectivamente.
    Por otra parte, los servicios de salud que registraron consumos con más frecuencia fueron los productos farmacéuticos y las consultas médicas, los cuales se verificaron en un 48,3 y 37,2 \% de los hogares, respectivamente. 
    En relación al \ac{GBS}, la media per cápita es de \$ 909, lo que representa una proporción media de 4,8 \% del gasto total.
    Por último, los indicadores de \ac{GCS} estimados muestran que un 15,6 \% de los hogares destinó más del 10 \% de su gasto total a \ac{GBS}, mientras que un 4,3 \% destinó más del 25 \% de su gasto.
    
    \begin{sidewaystable}[htp!]
        \centering
        \caption{Estadística descriptiva de variables utilizadas en los modelos \ac{RDD} y \ac{RPLP}}
        \label{tab:rdd.est.desc}
        \input{tablas/rdd.desc}
        \flushleft
        {\footnotesize\textit{Nota.} Ver descripción de las variables en la Tabla \ref{tab:rdd.vars}. Abreviaturas: n = tamaño de la muestra, PP = proporción, DE = desvío estándar, \ac{GCS} = gasto catastrófico en salud, \ac{PAMI} = Plan de Asistencia Médica Integral, JH = jefe de hogar. \\}
    \end{sidewaystable}

\begin{figure}[H]
        \vspace{5mm}
        \caption{Estimación de la proporción de jefes de hogar con distintos tipos de seguro según la distancia hasta la edad jubilatoria ($z=0$) en Argentina (2017/18).}
        \label{fig:rdd.seguro}
        \begin{subfigure}{.5\textwidth}
            \centering  
            \caption{JH con \ac{PAMI}.}\vspace{-2.5mm}
            \label{fig:rdd.pami}
            \includegraphics[width=.8\linewidth]{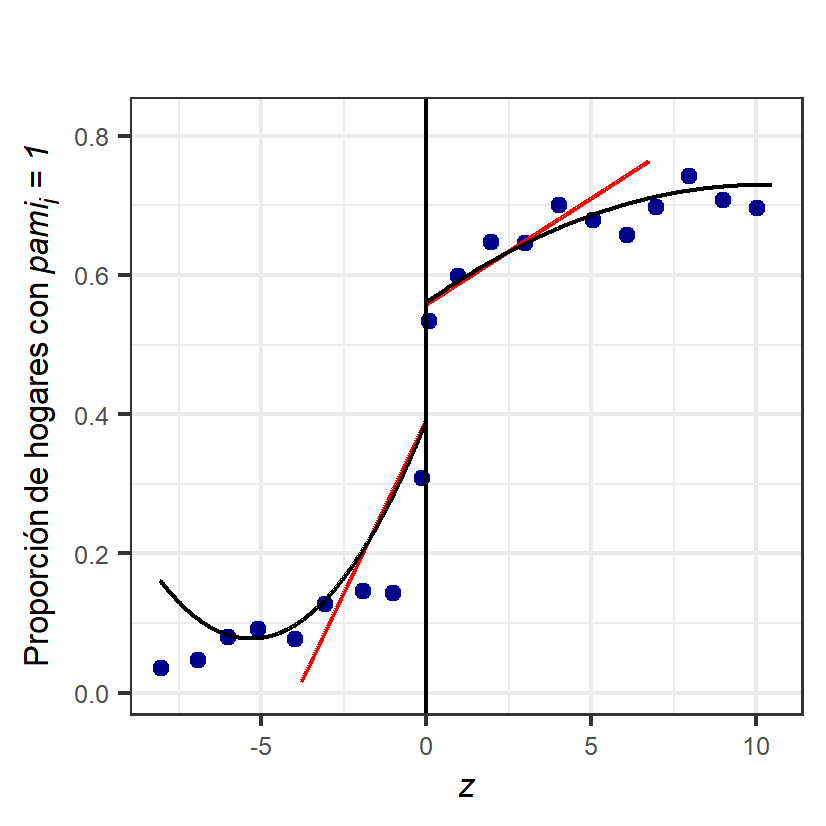}
        \end{subfigure}
        \begin{subfigure}{.5\textwidth}
            \centering
            \caption{JH con al menos un seguro.}\vspace{-2.5mm}
            \label{fig:rdd.seg}
            \includegraphics[width=.8\linewidth]{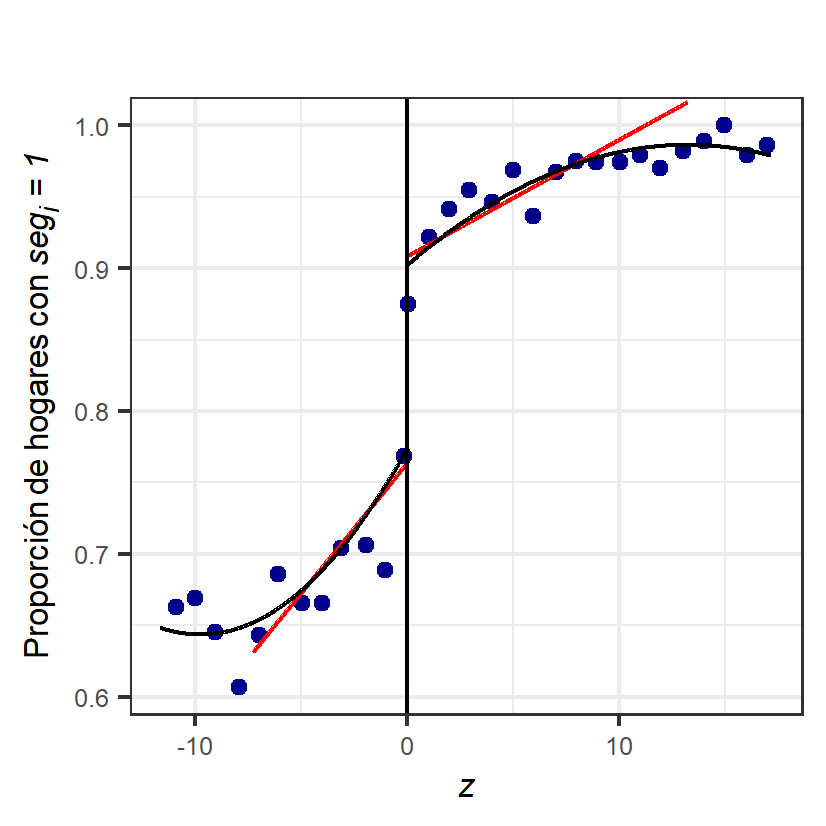}  
        \end{subfigure}
        \vfill
        \vspace{5mm}
        \begin{subfigure}{.5\textwidth}
            \centering
            \caption{JH con seguro voluntario voluntario}\vspace{-2.5mm}
            \label{fig:rdd.segp}
            \includegraphics[width=.8\linewidth]{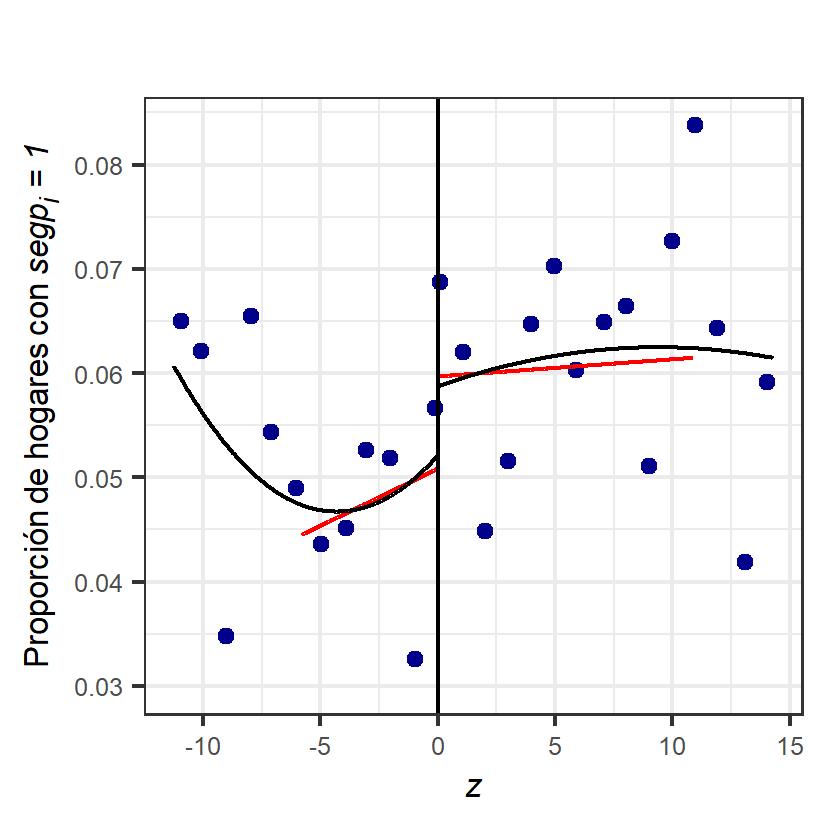}  
        \end{subfigure}
        \begin{subfigure}{.5\textwidth}
            \centering
            \caption{JH con más de un seguro.}\vspace{-2.5mm}
            \label{fig:rdd.segm}
            \includegraphics[width=.8\linewidth]{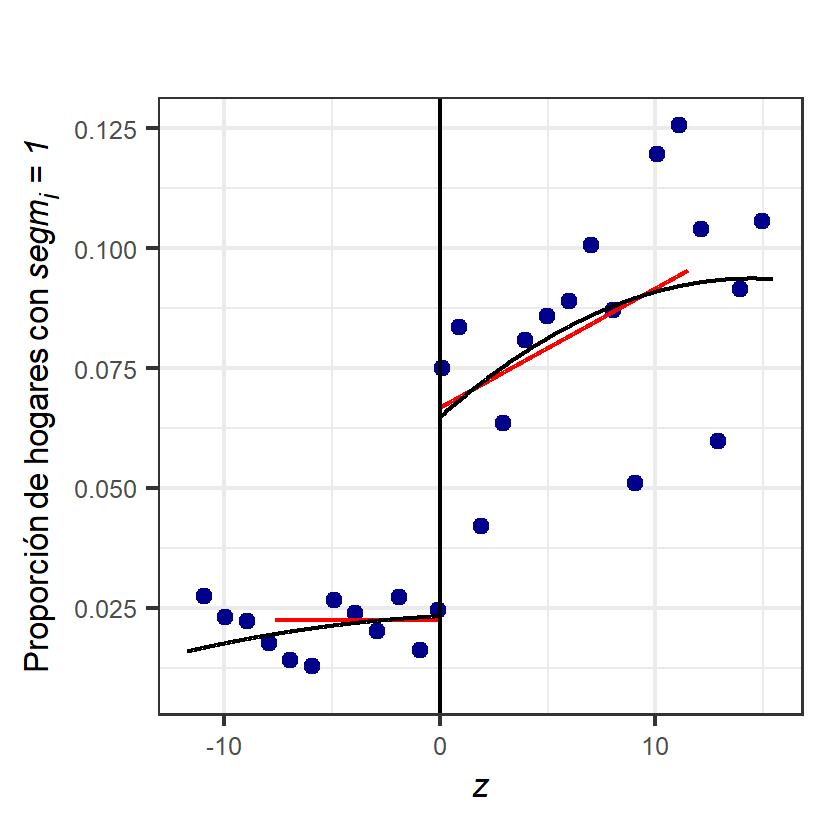}  
        \end{subfigure}
        \begin{subfigure}[c]{\textwidth}
            \centering
            \includegraphics{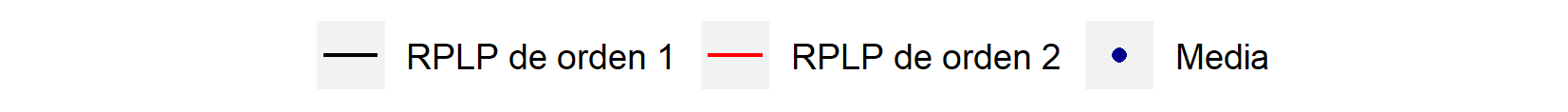}
        \end{subfigure}
        \flushleft
        {\footnotesize \textit{Nota.} Abreviaturas: JH = jefe de hogar, \ac{RPLP} = regresión polinómica local ponderada.
        Las \ac{RPLP} que se muestran en las figuras son estimados a la derecha e izquierda de $z=0$ utilizando los anchos de banda que se muestran en la Tabla \ref{tab:rdd.rkd.bws} y ponderadores Kernel triangulares (ver sección \ref{sec:rdd.datos.met}). La descripción de las variables que se miden en los ejes de las ordenadas se muestra en la Tabla \ref{tab:rdd.vars}. La variable $z_{i}$ es igual a la edad del JH menos 61 en caso que sea de género femenino y menos 66 si es de género masculino.}
    \end{figure}
\begin{figure}[H]
        \vspace{5mm}
        \caption{Estimación de indicadores de \ac{GBS} y \ac{GCS} para los hogares según la distancia de la edad del jefe de hogar hasta la edad jubilatoria ($z=0$) en Argentina (2017/18).}
        \label{fig:rdd.gasto}
        \begin{subfigure}{.5\textwidth}
            \centering  
            \caption{Log. del gasto en salud}\vspace{-2.5mm}
            \label{fig:rdd.lgsalud}
            \includegraphics[width=.8\linewidth]{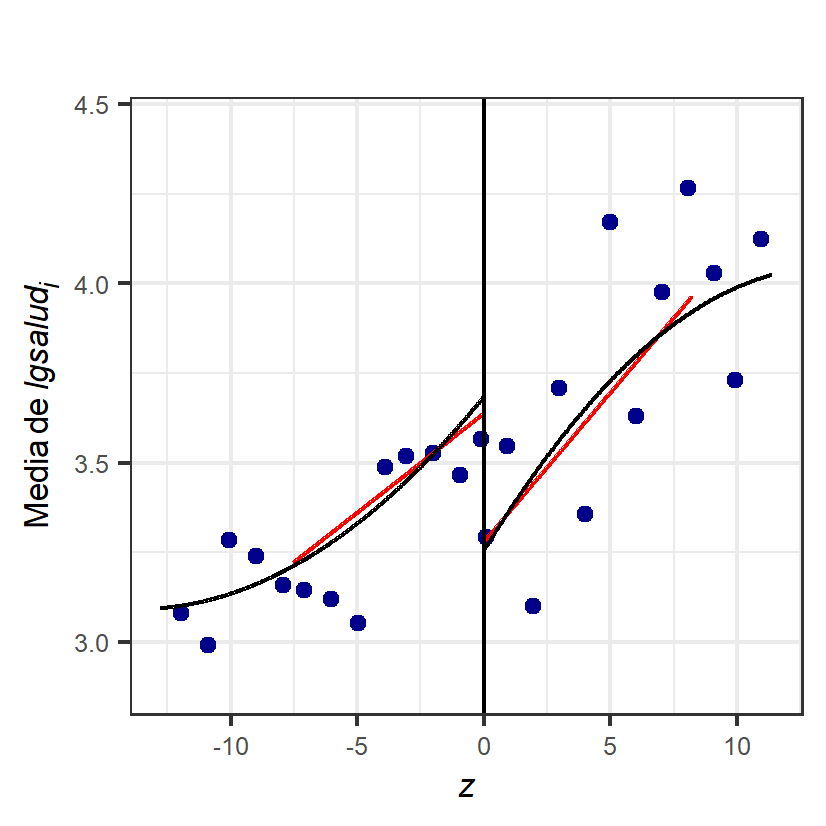}
        \end{subfigure}
        \begin{subfigure}{.5\textwidth}
            \centering
            \caption{Proporción del \ac{GBS} sobre el gasto total}\vspace{-2.5mm}
            \label{fig:rdd.gbs}
            \includegraphics[width=.8\linewidth]{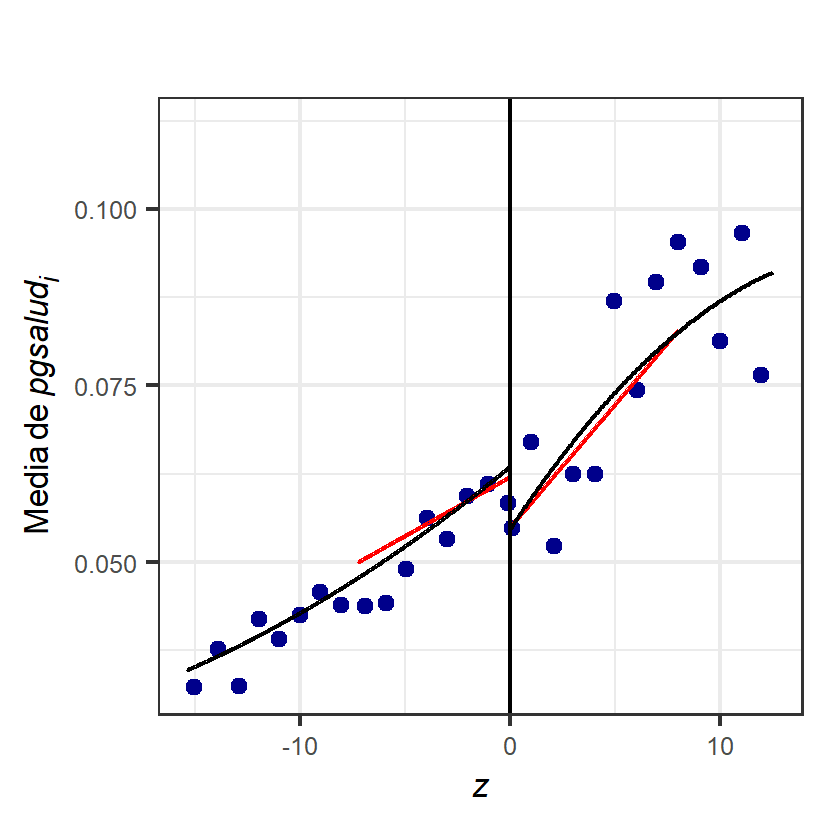}  
        \end{subfigure}
        \vfill
        \vspace{5mm}
        \begin{subfigure}{.5\textwidth}
            \centering
            \caption{\ac{GCS} ($cat^{10\%}$)}\vspace{-2.5mm}
            \label{fig:rdd.cat10}
            \includegraphics[width=.8\linewidth]{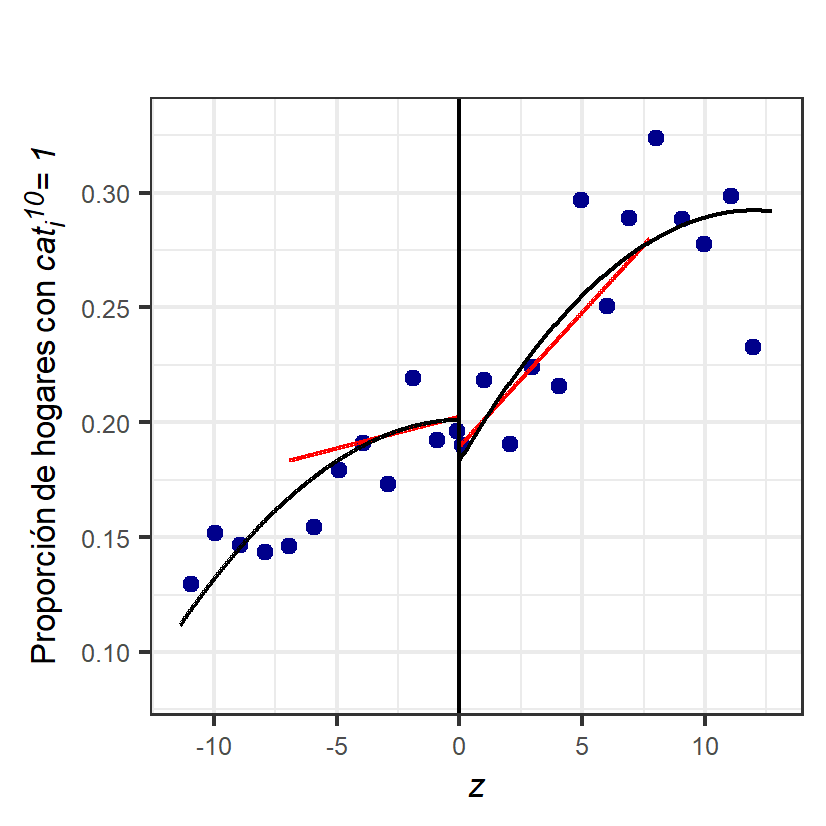}  
        \end{subfigure}
        \begin{subfigure}{.5\textwidth}
            \centering
            \caption{\ac{GCS} ($cat^{25\%}$)}\vspace{-2.5mm}
            \label{fig:rdd.cat25}
            \includegraphics[width=.8\linewidth]{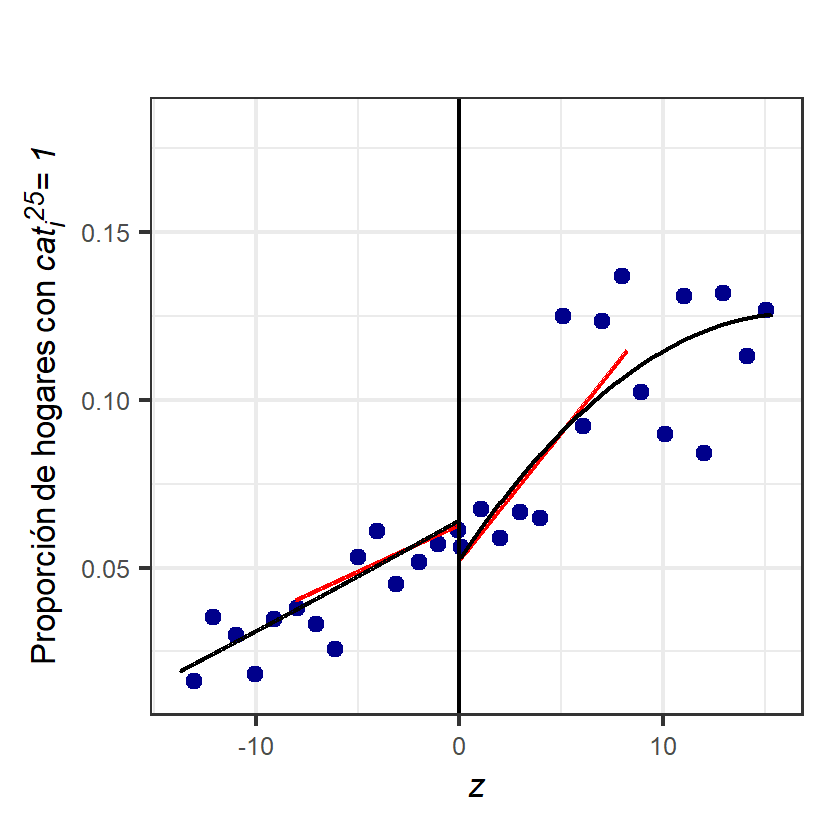}  
        \end{subfigure}
        \begin{subfigure}[c]{\textwidth}
            \centering
            \includegraphics{imagenes/rd.legend.png}
        \end{subfigure}
        \flushleft
        {\footnotesize \textit{Nota.} \ac{GBS} = gasto de bolsillo en salud, \ac{GCS} = gasto catastrófico en salud, \ac{RPLP} = regresión polinómica local ponderada.
        Las \ac{RPLP} que se muestran en las figuras son estimados a la derecha e izquierda de $z=0$ utilizando los anchos de banda que se muestran en la Tabla \ref{tab:rdd.rkd.bws} y ponderadores Kernel triangulares (ver sección \ref{sec:rdd.datos.met}). La descripción de las variables que se miden en los ejes de las ordenadas se muestra en la Tabla \ref{tab:rdd.vars}. La variable $z_{i}$ es igual a la edad del jefe de hogar menos 61 en caso que sea de género femenino y menos 66 si es de género masculino.}
    \end{figure}
\begin{figure}[H]
        \vspace{5mm}
        \caption{Proporciones de hogares que consumieron distintos bienes y servicios de salud según la distancia de la edad del jefe de hogar hasta la edad jubilatoria ($z=0$) en Argentina (2017/18).}
        \label{fig:rdd.bsyss}
        \begin{subfigure}{.5\textwidth}
            \centering  
            \caption{Servicios farmacéuticos}\vspace{-2.5mm}
            \label{fig:rdd.farm}
            \includegraphics[width=.8\linewidth]{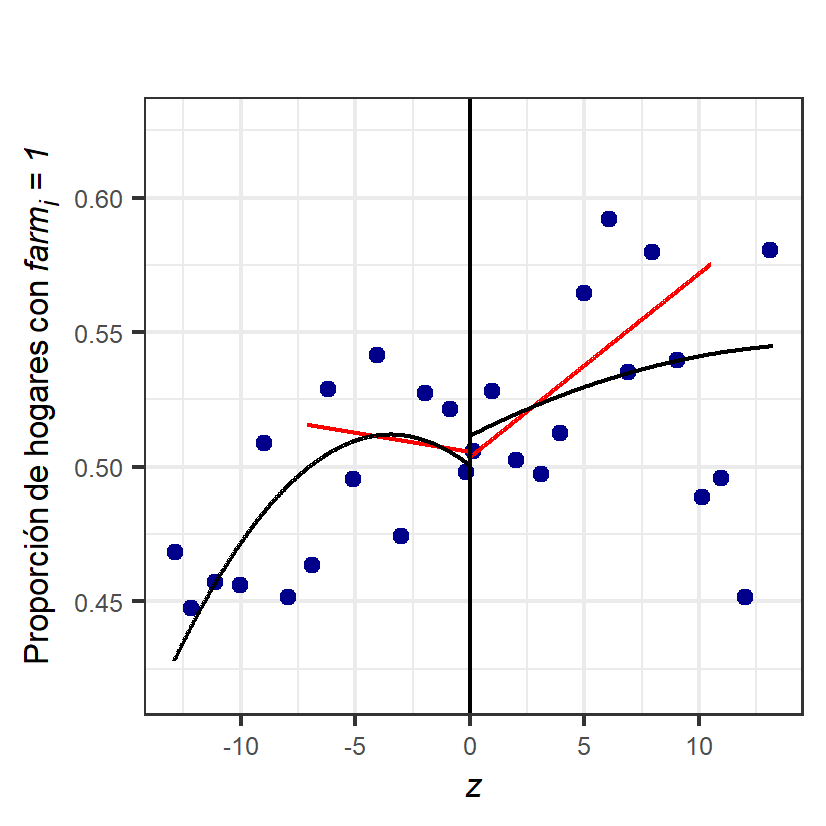}
        \end{subfigure}
        \begin{subfigure}{.5\textwidth}
            \centering
            \caption{Equipamiento}\vspace{-2.5mm}
            \label{fig:rdd.equi}
            \includegraphics[width=.8\linewidth]{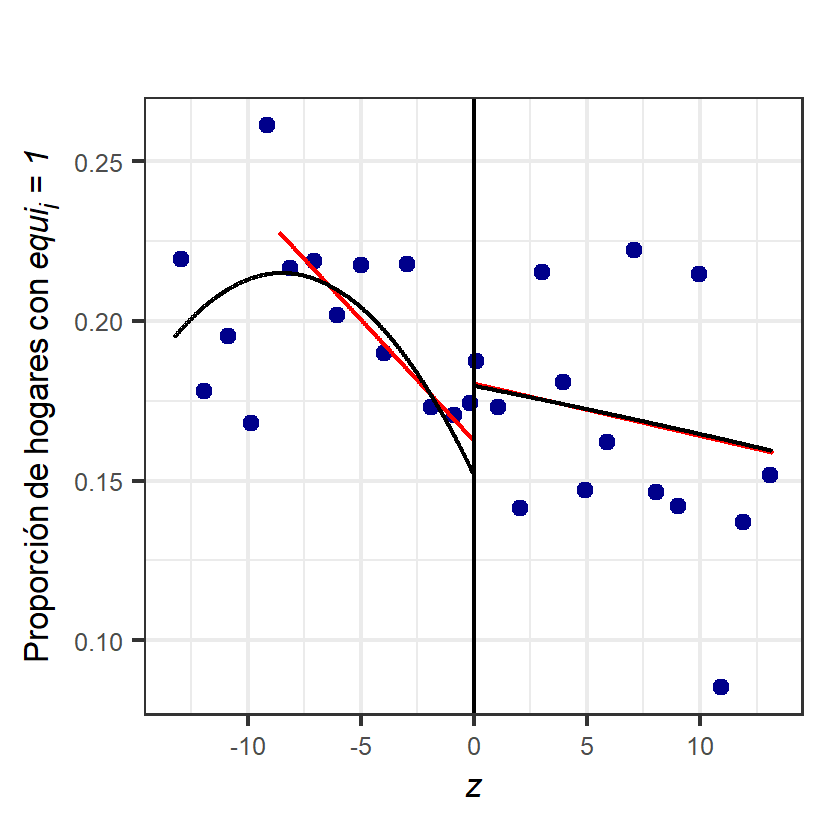}  
        \end{subfigure}
        \vfill
        \vspace{5mm}
        \begin{subfigure}{.5\textwidth}
            \centering
            \caption{Servicios médicos}\vspace{-2.5mm}
            \label{fig:rdd.smed}
            \includegraphics[width=.8\linewidth]{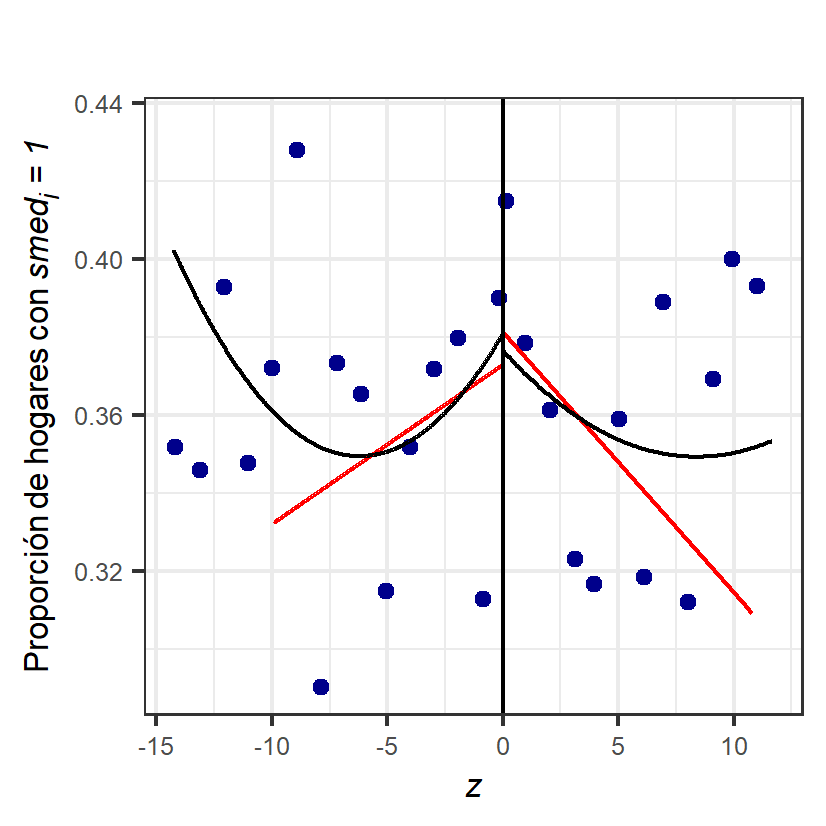}  
        \end{subfigure}
        \begin{subfigure}{.5\textwidth}
            \centering
            \caption{Servicios odontológicos}\vspace{-2.5mm}
            \label{fig:rdd.odon}
            \includegraphics[width=.8\linewidth]{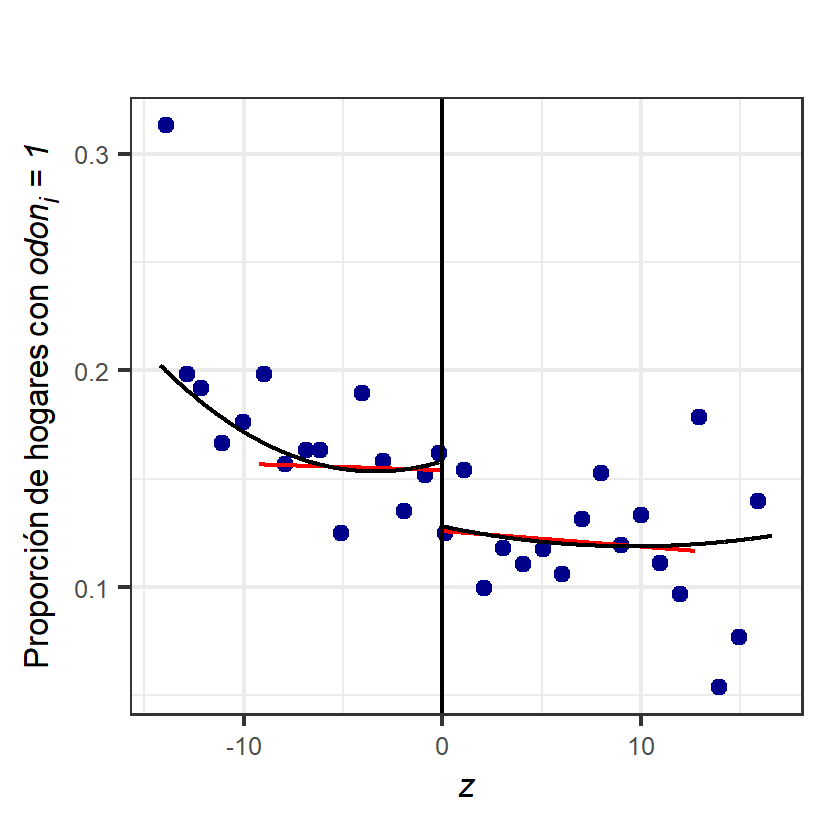}  
        \end{subfigure}
        \begin{subfigure}[c]{\textwidth}
            \centering
            \includegraphics{imagenes/rd.legend.png}
        \end{subfigure}
        {\footnotesize \textit{Nota.} \ac{GBS} = gasto de bolsillo en salud, \ac{GCS} = gasto catastrófico en salud, \ac{RPLP} = regresión polinómica local ponderada.
        Las \ac{RPLP} que se muestran en las figuras son estimados a la derecha e izquierda de $z=0$ utilizando los anchos de banda que se muestran en la Tabla \ref{tab:rdd.rkd.bws} y ponderadores Kernel triangulares (ver sección \ref{sec:rdd.datos.met}). La descripción de las variables que se miden en los ejes de las ordenadas se muestra en la Tabla \ref{tab:rdd.vars}. La variable $z_{i}$ es igual a la edad del jefe de hogar menos 61 en caso que sea de género femenino y menos 66 si es de género masculino.}
    \end{figure}

    Las Figuras \ref{fig:rdd.seguro}, \ref{fig:rdd.gasto} y \ref{fig:rdd.bsyss} que se presentan en esta sección muestran los valores esperados a la derecha e izquierda del punto de corte $z=0$ para la variable tratamiento y las variables objetivo.
    En los gráficos se muestra la media de estas variables para valores enteros de $z$, y las estimaciones a través de \ac{RPLP} de orden 1 y 2, cuya salto estimado en $z_{0}$ se presenta en la Tabla \ref{tab:rdd.resultados}.
    Los anchos de banda $h$ utilizados en cada estimación pueden consultarse en la Tabla \ref{tab:rdd.rkd.bws} del .
    Los resultados muestran que en $z=0$ se produce un cambio significativo en la variable tratamiento $pami_{i}$.
    La proporción de jefes de hogar beneficiarios del \ac{PAMI} se incrementa en 0,14 ($PV<0,001$), lo cual puede observarse en la Figura \ref{fig:rdd.seguro} (\subref{fig:rdd.pami}).
    Este incremento es de una magnitud importante, pues implica que el 14 \% de los jefes de hogar se incorporan al \ac{PAMI} en la edad jubilatoria.
    
    El impacto del \ac{PAMI} en la proporción de jefes de hogar que tienen al menos un seguro es de 0,5 ($PV<0,001$).
    Este resultado implica que si el 100 \% de la población accediera al \ac{PAMI} en la edad jubilatoria, el porcentaje de jefes de hogar que serían titulares de un seguro de salud se incrementaría en un 48 \%.
    Por otra parte, se observa que el \ac{PAMI} impacta en la cantidad de jefes de hogar con más de un seguro, con un $\hat{\tau}$ de 0,16 en la estimación de orden 1 y 0,21 en la estimación de orden 2 ($PV<0,001$ en todos los casos).
    El valor $\hat{\tau}$ para el seguro voluntario fue cercano a cero sin significatividad estadística, lo que indica que no se verificó un efecto \textit{crowding out} sobre el mercado voluntario de seguros de salud.
    Los efectos descriptos pueden observarse en la Figura \ref{fig:rdd.seguro}, la cual muestran saltos de los valores medios en $z_{0}$ para todos los tipos de seguro del jefe de hogar, a excepción del seguro voluntario.
    
    En relación a la protección financiera, los resultados basados en las  \ac{RPLP} que se presentan en la Figura \ref{fig:rdd.gasto}  muestran un cambio negativo en el punto de corte $z=0$.
    En el caso del gasto en salud per cápita, este disminuye cerca del 2 \% ($PV<0,01$ en la estimación de orden 1 y $PV<0,05$ en la de orden 2), mientras que la proporción del gasto en salud disminuye un 4 \% ($PV<0,1$).
    Los indicadores de \ac{GCS} mostraron parámetros estimados, que señalan caídas de entre el 0,06 y el 0,11 en la proporción de hogares, pero sin significatividad estadística ($PV>0,1$).
    
    Los resultados relacionados con el acceso a bienes y servicios de salud muestran parámetros con signos opuestos al esperado o sin significatividad estadística.
    La proporción de hogares que realizó consumos de servicios farmacéuticos tiene una variación en el punto de corte de -0,04 ($PV>0,1$) en la estimación de orden 1, y de 0,01 ($PV>0,1$) en la estimación de orden 2. 
    Por otra parte, las estimaciones para servicios médicos y equipamiento tienen parámetros con el signo esperado, pero sin significatividad estadística.
    Finalmente, los servicios odontológicos muestran una discontinuidad de -0,13 a partir de la \ac{RPLP} de orden 1 y de -0,17 en la de orden 2, pero con parámetros no significativos.
    Estos resultados son consistentes con la dispersión de puntos que se presenta en la Figura \ref{fig:rdd.bsyss}, donde se observan nubes sin discontinuidades claras.
    
    Los parámetros $\hat{\tau}$ estimados en los \ac{RKD} se presentan en la Tabla \ref{tab:rkd.resultados} y los anchos de banda $h$ utilizados en la Tabla \ref{tab:rdd.rkd.bws}, las cuales pueden econtrarse en el anexo.
    Las estimaciones no arrojaron resultados significativos, lo que indica que no hay evidencia sobre un cambio en la tasa de variación de las variables objetivo alrededor de la edad jubilatoria $z=0$.
    
    \begin{table}[H]
        \centering
        \caption{\ac{RDD} de distintas variables en la edad jubilatoria ($z=0$).}
        \label{tab:rdd.resultados}
        \input{tablas/rdd.resultados}
        \flushleft
        {\footnotesize\textit{Nota.} $^{***}PV<0,001$, $^{**}PV<0,01$, $^{*}PV<0.05$, $^{\bullet} PV<0.1$. \\
        Abreviaturas: $p$ = orden de la regresión, DE = desvío estándar, \ac{GBS} = gastos de bolsillo en salud, N = tamaño de la muestra (se presenta el número de observaciones utilizado a izquierda y derecha del punto de corte).\\
        Ver descripción de las variables en la Tabla \ref{tab:rdd.vars}. \\
        \textsuperscript{1}\ac{RDD} \textit{sharp} de la variable objetivo en $z=0$. \\
        \textsuperscript{2}\ac{RDD} \textit{fuzzy} de la variable objetivo en $z=0$, utilizando como tratamiento $pami_{i}$. \\
        \textsuperscript{3}Se utilizaron como covariables el logarítmo del gasto per cápita del hogar ($lgasto_{i}$), el sexo del jefe de hogar ($sexo_{i}$) y la inactividad económica del jefe de hogar ($inac_{i}$).
        }
    \end{table}

\section{Conclusiones y discusión}

    El \ac{PAMI} es una de las políticas públicas más importantes dentro del \ac{SSA} y la más importante para la financiación de la salud de los adultos mayores en Argentina.
    En este capítulo se encontró un efecto importante del \ac{PAMI} en la cantidad de jefes de hogar que tienen seguro de salud.
    Este resultado es relevante para todos los integrantes de los hogares dado que la reglamentación del \ac{PAMI} permite que un beneficiario titular incorpore a su grupo familiar dentro del seguro de salud.
    Sin embargo, los resultados sobre el efecto en la cobertura en salud son menos alentadores. 
    Las evaluaciones realizadas sobre el \ac{GBS} mostraron mejoras pequeñas en la protección financiera, y no se encontraron resultados a partir de los indicadores de \ac{GCS}.
    Por otra parte, no se encontraron mejoras en el acceso a la salud medida en términos de la utilización de servicios.
    
    El estudio presentado en este capítulo es similar al realizado por \textcite{Card2008-vv} para el análisis del programa \ac{MEDICARE} en Estados Unidos, en el cual evalúan su efecto en la utilización de servicios médicos.
    Al igual que en el caso del \ac{PAMI} en la edad jubilatoria, encuentran que a los 65 años se produce un salto en la proporción de la población que tiene un seguro de salud y, a partir de este hecho, evalúan la existencia de saltos en otras variables.
    Los autores encuentran un efecto similar sobre la posesión de más de un seguro como el que se observa en la Figura \ref{fig:rdd.seguro} (\subref{fig:rdd.seg}), pero, a diferencia de los resultados presentados en este capítulo, también encuentran cambios en la utilización de servicios.

    Una posible explicación de estos resultados reside en la heterogeneidad de las características de la población que accede al \ac{PAMI}.
    En primer lugar, existe un grupo de la población beneficiaria de un seguro a través de una obra social o de un seguro voluntario en su etapa activa.
    En segundo lugar, como se observa en la Figura \ref{fig:rdd.seguro} (\subref{fig:rdd.seg}), existe un grupo de jefes de hogar, que representaría entre el 15 y el 20 \% del total, que llegan a la edad jubilatoria sin seguro.
    En caso que este grupo, al convertirse en beneficiario del \ac{PAMI}, no modifique sus pautas de utilización del sistema de salud, podría verificarse un salto en la proporción de población con seguro de salud, manteniéndose constante la utilización de servicios y la protección financiera.
    Por lo tanto, los resultados observados podrían explicarse por nuevos beneficiarios del \ac{PAMI} que no tenían seguro en su etapa activa, y continúan accediendo a prestaciones sanitarias a través de \ac{GBS} o utilizando la oferta del subsistema público.
Este resultado es consistente con el comportamiento de los jefes de hogar relacionados a la titularidad de más de un seguro.
En primer lugar, en la Figura \ref{fig:rdd.seguro} (\subref{fig:rdd.segm}) se observa que una proporción cercana al 5 \% decide mantener un seguro adicional.
En segundo lugar, la ausencia de efecto \textit{crowding-out} muestra que los nuevos beneficiarios del \ac{PAMI} están dispuestos a conservar un seguro voluntario, es decir que no ven al \ac{PAMI} como un sustituto perfecto de otros tipos de seguro de salud. 
Esto es confirmado por las regresiones estimadas a la derecha del punto de corte, las cuales tienen una leve pendiente positiva.

    Otra posible explicación para la ausencia de cambios en la utilización de servicios podría encontrarse en que, a partir de que una persona sin seguro se convierte en beneficiario de \ac{PAMI}, transita un periodo de adaptación que imposibilita la detección de un salto en las variables objetivo.
    De todas formas, se debe considerar que, en general, las prestaciones médicas son costosas y existe un 27 \% de jefes de hogar que no cuentan con seguro de salud. 
    De no existir el sistema público, parecería poco probable la inexistencia de una demanda contenida en personas de entre 60 y 65 años, la cual debería efectivizarse al momento de una caída del costo de las prestaciones a valores cercanos a cero.

    La verificación de las hipótesis mencionadas previamente tienen las implicancias que se mencionan a continuación. 
    Si se corrobora la existencia de beneficiarios que estén abonando \ac{GBS} por prestaciones a las cuales deberían acceder sin costo, el \ac{PAMI} debería revisar su oferta de servicios, generando auditorías para detectar pagos informales. 
    En caso que se verifique que los beneficiarios están incurriendo en \ac{GBS} para acceder a prestaciones que no están incluidas dentro de las ofrecidas por \ac{PAMI}, este organismo deberá evaluar la factibilidad y el beneficio de su inclusión.
    Por último, si se verifica que los afiliados acuden al subsistema público, el \ac{PAMI} junto con el Estado deberían organizar su oferta prestacional de forma que los afiliados sean solo recibidos en prestadores que han firmado un convenio con \ac{PAMI}, con el objetivo de evitar subsidios cruzados entre el Estado y el \ac{PAMI}.
    
Los resultados de este capítulo deben evaluarse considerando las siguientes limitaciones. En primer lugar, en la \ac{ENGHo} la unidad observacional es el hogar, lo que dificulta efectuar estimaciones que incluyan a la edad o el seguro de salud en hogares heterogéneos. 
La caracterización a partir del jefe de hogar da lugar a distintas imprecisiones. 
Por ejemplo, un hogar en el que todos sus integrantes tienen cobertura a excepción del jefe de hogar, y en el que se verifica utilización de bienes y servicios relacionados con la salud sería erróneamente clasificado como un hogar sin seguro de salud que accedió al sistema. 
Esta limitación afecta a la precisión de los resultados relacionados al \ac{GBS} al acceso a servicios de salud. 
En segundo lugar, el tamaño de la muestra para los datos de utilización de servicios de salud es sensiblemente menor a las variables de gasto y de posesión de seguro. 
Como consecuencia, la potencia de los test estadísticos es menor, lo cual debilita la aceptación de la hipótesis de continuidad de las variables en el punto de corte.
Por último, la forma de medir acceso a la salud está basada en la utilización de un conjunto servicios expresados como variables dicotómicas, lo cual no incluye a otros servicios que pueden resultar relevantes y no permite evaluar diferencias entre las cantidades utilizadas.

Los resultados encontrados muestran la importancia de estudiar las barreras en el acceso a servicios de salud que enfrentan los beneficiarios del \ac{PAMI}, las cuales pueden entenderse como los obstáculos para que se cumplan las condiciones de acceso a la salud.
El cambio en la población asegurada es de una magnitud sobre la cual parece inverosímil que la utilización de servicios no se modifique.
En este sentido, existen, al menos, cuatro dimensiones que es necesario explorar.
En primer lugar, es necesario conocer como influyen los usos y costumbres en el acceso a la salud, lo cual incluye una evaluación de cambios lentos en los hábitos de consumo a partir de la obtención de un seguro de salud. 
Dentro de la población que se incorpora al \ac{PAMI}, existe una proporción que no tiene seguro antes de obtenerlo, por lo cual no tiene experiencia en trámites administrativos específicos de los seguros de salud y este factor puede constituir una barrera en el acceso.
En segundo lugar, podrían existir barreras causadas por limitaciones en la oferta de servicios.
Por último, se debe evaluar la diferencia entre la oferta del subsistema público y los servicios que tienen disponibles los beneficiarios del \ac{PAMI} dado que, en caso de ser similares, no se observarían diferencias en la utilización de servicios.

\section{Referencias bibliográficas}
\printbibliography[heading=none]

\appendix
\pagebreak
\section{Apéndice}

   \begin{table}[htp!]
    \small
        \centering
        \caption{\ac{RKD} de distintas variables en la edad jubilatoria ($z=0$).}
        \label{tab:rkd.resultados}
        \input{tablas/rkd.resultados}
        \flushleft
        {\footnotesize\textit{Nota.}  Ver descripción de las variables en la Tabla \ref{tab:rdd.vars}.
        Abreviaturas: $p$ = orden de la regresión, DE = desvío estándar, \ac{GBS} = gastos de bolsillo en salud.\\
        \textsuperscript{1}\ac{RKD} \textit{sharp} de la variable objetivo en $z=0$. \\
        \textsuperscript{2}\ac{RKD} \textit{fuzzy} de la variable objetivo en $z=0$, utilizando como tratamiento $pami_{i}$. \\
        \textsuperscript{3}Se utilizaron como covariables el logarítmo del gasto per cápita del hogar ($lgasto_{i}$), el sexo del jefe de hogar ($sexo_{i}$) y la inactividad económica del jefe de hogar ($inac_{i}$).\\
        $^{***}PV<0,001$, $^{**}PV<0,01$, $^{*}PV<0.05$, $^{\bullet} PV<0.1$.
        }
    \end{table}

\begin{table}[htp!]
\footnotesize
    \centering
    \caption{Anchos de banda $h$ utilizados en los \ac{RDD} presentados en la Tabla \ref{tab:rdd.resultados} y en los \ac{RKD} presentados en la Tabla \ref{tab:rkd.resultados}}
    \label{tab:rdd.rkd.bws}
    \input{tablas/rdd.rkd.bws}
        \flushleft
        {\footnotesize\textit{Nota.} Ver descripción de las variables en la Tabla \ref{tab:rdd.vars}.
        Abreviaturas: $p$ = orden de la regresión, \ac{GBS} = gastos de bolsillo en salud, BWL = ancho de banda a la izquierda de $z=0$, BWR = ancho de banda a la derecha de $z=0$.\\
        \textsuperscript{1}\ac{RKD} \textit{sharp} de la variable objetivo en $z=0$. \\
        \textsuperscript{2}\ac{RKD} \textit{fuzzy} de la variable objetivo en $z=0$, utilizando como tratamiento $pami_{i}$. \\
        \textsuperscript{3}Se utilizaron como covariables el logarítmo del gasto per cápita del hogar ($lgasto_{i}$), el sexo del jefe de hogar ($sexo_{i}$) y la inactividad económica del jefe de hogar ($inac_{i}$).\\
        $^{***}PV<0,001$, $^{**}PV<0,01$, $^{*}PV<0.05$, $^{\bullet} PV<0.1$.
        }
\end{table}

\end{document}

%% file: bibliography.tex
\usepackage[citestyle=authoryear, alldates=year]{biblatex}
\DeclareDelimFormat{finalnamedelim}{\addspace\bibstring{and}\space}

\DefineBibliographyStrings{english}{
  and = {y}
}

\defbibenvironment{bibliography}
{\list
{\printfield[labelnumberwidth]{}}
{\setlength{\leftmargin}{\bibhang}%
\setlength{\itemindent}{-\leftmargin}%
\setlength{\itemsep}{\bibitemsep}%
\setlength{\parsep}{4mm}}}
{\endlist}
{\item}

\DeclareExtradate{%
    \scope{
        \field{sortkey}
    }
}

\DeclareNameFormat{author}{
    \ifthenelse{\value{listcount}=1}
    {\namepartfamily\space%
    \namepartgiveni}{%
        \ifthenelse{\value{listcount}<\value{liststop}}%
        {\addcomma\space\namepartfamily\space%
        \namepartgiveni}{%
            \ifthenelse{\value{listcount}=\value{liststop}}{%
            \&\space\namepartfamily\space%
            \namepartgiveni}{}}}
}

\DeclareFieldFormat*{title}{#1}
\DeclareFieldFormat*{pages}{#1}
\DeclareFieldFormat*{italics}{\textit{#1}}
\DeclareFieldFormat*{publisher}{#1}

\newbibmacro*{journ+vol+num}{%
    \printfield{journaltitle}%
    \iffieldundef{volume}{}{%
        \iffieldundef{number}%
        {, \printfield[italics]{volume}}%
        {, \printfield[italics]{volume}(\printfield{number})}}%
    \iffieldundef{pages}{}{, \printfield{pages}}
}

\DeclareBibliographyDriver{article}{%
    \printnames{author}%
    (\printdateextra)%
    \newunit
    \printfield{title}%
    \newunit\newblock
    \usebibmacro{journ+vol+num}%
    \newunit
    \iffieldundef{doi}{\printfield{url}}{\printfield{doi}}
    \finentry
}

\DeclareBibliographyDriver{book}{%
    \printnames{author}%
    (\printdateextra)%
    \newunit
    \printfield[italics]{title}%
    \iffieldundef{edition}{}{( \printfield{edition} ed.)}%
    \newunit
    \printlist{publisher}
    \newunit
    \iffieldundef{doi}{\printfield{url}}{\printfield{doi}}
    \finentry
}

\DeclareBibliographyDriver{misc}{%
    \printnames{author}%
    (\printdateextra)%
    \newunit
    \printfield{title}%
    \newunit
    \printlist{publisher}
    \newunit
    \iffieldundef{doi}{\printfield{url}}{\printfield{doi}}
    \finentry
}

%% file: tablas/rdd.vars.tex
	\begin{tabular}{p{.1\textwidth}p{.85\textwidth}}
		\hline\hline \\[-0ex]
		\multicolumn{2}{c}{\quad Variables objetivo}  \\\hline
		$lgsalud_{i}$ & Logaritmo natural del gasto per cápita en salud del hogar. \\
		$pgsalud_{i}$ & Proporción de gasto en salud respecto al gasto total del hogar. \\
		$cat^{10\%}$ & Hogar que destinó más del 10 \% de su gasto total a \ac{GBS} (Sí = 1; No = 0). \\
		$cat^{25 \%}$ & Hogar que destinó más del 25 \% de su gasto total a \ac{GBS} (Sí = 1; No = 0). \\
		$seg_{i}$ & El jefe de hogar es titular de un seguro de salud (Sí = 1; No = 0).\\
		$segp_{i}$ & El jefe de hogar es titular de un seguro voluntario (Sí = 1; No = 0). \\
		$segm_{i}$ & El jefe de hogar es titular de dos seguros o más (Sí = 1; No = 0). \\
		$farm_{i}$ & Algún integrante del hogar adquirió productos farmacéuticos (Sí = 1; No = 0).\\
		$equi_{i}$ & Algún integrante del hogar adquirió artefactos o equipos terapéuticos (Sí = 1; No = 0).\\
		$smed_{i}$ & Algún integrante del hogar realizó una consulta médica (Sí = 1; No = 0).\\
		$odon_{i}$ & Algún integrante del hogar recibió servicios odontológicos (Sí = 1; No = 0).\\
		$hosp_{i}$ & Algún integrante del hogar recibió servicios hospitalarios (Sí = 1; No = 0).\\ & \\[-0ex]
		\multicolumn{2}{c}{\quad Variable tratamiento} \\\hline
		$pami_{i}$ & El jefe de hogar es beneficiario del PAMI (Sí = 1; No = 0).\\ & \\[-0ex]
		\multicolumn{2}{c}{Variable observable que asigna el tratamiento}  \\\hline
		$edad_{i}$ & Edad del jefe de hogar. \\ &  \\[-0ex]
		\multicolumn{2}{c}{\quad Covariables} \\\hline
		$lgasto_{i}$ & Logaritmo natural del gasto per cápita del hogar. \\
		$sexo_{i}$ & Sexo del jefe de hogar (Mujer = 1; Varon = 0). \\
		$inac_{i}$ & Jefe de hogar económicamente inactivo (Sí = 1; No = 0). \\
		\\\hline\hline
	\end{tabular}

%% file: tablas/rdd.desc.tex
\begin{tabular}{@{\extracolsep{5pt}}llcccccccc} 
\\[-1.8ex]\hline 
\hline 

\multicolumn{2}{c}{} & \multicolumn{4}{c}{Grupo tratamiento} & \multicolumn{4}{c}{Grupo control} \\\cline{3-6}\cline{7-10} 

\multicolumn{1}{c}{Variable} & \multicolumn{1}{c}{Notación} & \multicolumn{1}{c}{n} & \multicolumn{1}{c}{Media} & \multicolumn{1}{c}{PP} & \multicolumn{1}{c}{DE}
 & \multicolumn{1}{c}{n} & \multicolumn{1}{c}{Media} & \multicolumn{1}{c}{PP} & \multicolumn{1}{c}{DE}\\ 
\hline \\[-1ex] 

\multicolumn{5}{l}{\quad \textit{Variables objetivo}} \\\hline
Gasto en salud & $gsalud_{i}$                           & 16.161 & 1.145,85 && 3.080,50 & 5.372 & 1.799,04 && 3.863,26 \\ 
Proporción del gasto en salud & $pgsalud_{i}$           & 16.161 && 0,04 & 0,07 & 5.372 && 0,08 & 0,13 \\ 
\ac{GCS}\textsuperscript{10\%} & $cat^{10\%}$           & 16.161 && 0,12 & 0,33 & 5.372 && 0,26 & 0,44 \\ 
\ac{GCS}\textsuperscript{25\%} & $cat^{25\%}$           & 16.161 && 0,02 & 0,15 & 5.372 && 0,10 & 0,30 \\ 
Posesión de un seguro de salud o más & $seg_{i}$        & 16.144 && 0,65 & 0,48 & 5.368 && 0,96 & 0,20 \\ 
Posesión de más de un seguro & $segm_{i}$               & 16.144 && 0,01 & 0,12 & 5.368 && 0,09 & 0,28 \\
Posesión de seguro voluntario & $segp_{i}$              & 16.144 && 0,04 & 0,20 & 5.368 && 0,06 & 0,24 \\
Utilización de servicios farmacéuticos & $farm_{i}$     & 8.588  && 0,46 & 0,50 & 3.242 && 0,54 & 0,50 \\ 
Adquisición de equipamiento médico  & $equi_{i}$        & 8.588  && 0,19 & 0,39 & 3.242 && 0,16 & 0,37 \\ 
Realización de consultas médicas & $smed_{i}$           & 8.588  && 0,38 & 0,48 & 3.242 && 0,36 & 0,48 \\ 
Utilización de servicios odontológicos & $sden_{i}$     & 8.588  && 0,18 & 0,38 & 3.242 && 0,12 & 0,32 \\ 
 & &&&& \\ [-0ex] 
  \multicolumn{5}{l}{\quad \textit{Variable tratamiento}} \\\hline 
\ac{PAMI} como seguro médico & $pami_{i}$                   & 16.144 && 0,04 & 0,18 & 5.368 && 0,70 & 0,46 \\  
 & &&&& \\ [-0ex] 
  \multicolumn{5}{l}{\quad \textit{Variable continua observada}} \\\hline 
Edad del JH menos edad jubilatoria & $edadn_{i}$ & 16.161 & $-$20,28 && 11,47 & 5.372 & 9,36 && 7,39 \\  
 & &&&& \\ [-0ex]

  \multicolumn{5}{l}{\quad \textit{Covariables}} \\\hline
Gasto total & $gasto_{i}$                                   & 16.161 & 23.792,45 && 20.047,89 & 5.372 & 17.801,32 && 15.521,00 \\ 
Sexo femenino & $sexo_{i}$                                  & 16.161 && 0,40 & 0,49 & 5.372 && 0,58 & 0,49 \\ 
Inactivo económicamente & $inac_{i}$                        & 16.161 && 0,15 & 0,36 & 5.372 && 0,82 & 0,38 \\  
 & &&& \\ [-0ex] 
\hline\hline \\[-1.8ex] 
\end{tabular} 

%% file: tablas/rdd.resultados.tex
\footnotesize

\begin{tabular}{lrllllll}
  \hline\hline
  & & \multicolumn{3}{c}{Con covariables\textsuperscript{3}} & \multicolumn{3}{c}{Sin covariables} \\\cline{3-5}\cline{6-8}
Variable objetivo & $p$ & \multicolumn{1}{c}{$\hat{\tau}$} & \multicolumn{1}{c}{DE} & \multicolumn{1}{c}{N\{Der.; Izq.\}} & \multicolumn{1}{c}{$\hat{\tau}$} & \multicolumn{1}{c}{DE} & \multicolumn{1}{c}{N\{Der.; Izq.\}}  \\ 
  \hline  &&&&&&\\
    \quad Variable tratamiento\textsuperscript{1} &&&&&& \\\hline &&&&&& \\[-1ex]
    $pami_i$    &   1   &   0.14*** &   0.04    &   \{1180; 2320\}  &   0.14*** &   0.04    & \{1180; 2320\}  \\ 
                &   2   &   0.14*** &   0.05    &   \{3180; 3377\} & 0.13*** &   0.05       & \{3180; 3377\}     \\[1.8ex]
  \quad Tipos de seguro\textsuperscript{2} &&&& \\\hline &&&& \\[-1ex]

    $seg_i$     &   1   &   0.48*** &   0.08    & \{2757; 3937\}  &   0.51*** &    0.08     & \{2757; 3377\} \\ 
   &   2 & 0.5*** & 0.12                        & \{4415; 4551\}  & 0.51*** & 0.12          & \{4815; 4265\}   \\[1.8ex]
  $segp_i$ &   1 & 0.02  & 0.07                 & \{1956; 3377\} & 0.04  & 0.05             & \{2757; 3770\} \\ 
   &   2 & 0.04  & 0.08                         & \{4415; 4123\} & 0.06  & 0.07             & \{5646; 3937\} \\ [1.8ex]
  $segm_i$ &   1 & 0.16*** & 0.04               & \{2757; 3568\} & 0.14*** & 0.04 & \{3980; 3770\} \\ 
   &   2 & 0.21** & 0.07                       & \{4415; 4265\} & 0.2*** & 0.06 & \{5982; 3937\} \\ [1.8ex]
 \multicolumn{3}{c}{\quad Acceso a bienes y servicios de salud\textsuperscript{2}} &&&& \\\hline &&&&&&& \\[-1ex]
  $farm_i$ &   1 & -0.04  & 0.18 & \{1562; 1989\} & -0.02  & 0.17 & \{1562; 2106\}  \\ 
  &   2 & 0.01  & 0.25 & \{2693; 2342\}  & 0.02  & 0.25 & \{2912; 2230\}  \\ [1.8ex]
  $equi_i$ &   1 & 0.08  & 0.11 & \{1794; 2342\}  & 0.09  & 0.11 & \{1794; 2230\}  \\ 
   &   2 & 0.11  & 0.19 & \{2912; 2230\}  & 0.13  & 0.18 & \{3149; 2230\}  \\ [1.8ex]
  $smed_i$ &   1 & 0.05  & 0.13 & \{2011; 1989\} & 0.07  & 0.13 & \{2233; 1989\} \\ 
   &   2 & 0.16  & 0.23 & \{3149; 2106\}  & 0.17  & 0.22 & \{3331; 2106\}  \\ [1.8ex]
  $odon_i$ &   1 & -0.13  & 0.11 & \{2011; 2230\}  & -0.14  & 0.12 & \{1562; 2230\}  \\ 
   &   2 & -0.17  & 0.16 & \{3149; 2637\}  & -0.17  & 0.16 & \{2912; 2637\}  \\ [1.8ex]

\quad \ac{GBS}\textsuperscript{2} &&&& \\\hline &&&& \\[-1ex]
  $lgsalud_i$ &   1 & -2.1** & 0.73 & \{2757; 2869\}  & -1.68. & 0.78 & \{2757; 3377\}  \\ 
   &   2 & -2.55* & 1.04 & \{4815; 3568\}  & -1.89. & 1.1 & \{5211; 3377\}  \\ [1.8ex]
  $pgsalud_i$ &   1 & -0.04$^{\bullet}$ & 0.02 & \{2757; 2628\}  & -0.04. & 0.02 & \{3577; 2869\}  \\ 
   &   2 & -0.04$^{\bullet}$ & 0.03 & \{5982; 3770\}  & -0.02  & 0.03 & \{6864; 3377\}  \\ [1.8ex]
  $cat10_i$ &   1 & -0.11  & 0.1 & \{2369; 2628\}  & -0.11  & 0.08 & \{3180; 3143\}  \\ 
   &   2 & -0.09  & 0.14 & \{4415; 3770\}  & -0.04  & 0.15 & \{4415; 3568\}  \\ [1.8ex]
  $cat25_i$ &   1 & -0.06  & 0.05 & \{3180; 2869\}  & -0.06  & 0.04 & \{5211; 3143\}  \\ 
   &   2 & -0.07  & 0.07 & \{5211; 4265\}  & -0.06  & 0.06 & \{6463; 4265\}  \\ [1.8ex]
   \hline\hline
\end{tabular}

%% file: tablas/rkd.resultados.tex
\begin{tabular}{lrllll}
  \hline\hline
  & & \multicolumn{2}{c}{Con covariables\textsuperscript{3}} & \multicolumn{2}{c}{Sin covariables} \\\cline{3-6}
Variable objetivo & $p$ & \multicolumn{1}{c}{$\hat{\tau}$} & \multicolumn{1}{c}{DE} & \multicolumn{1}{c}{$\hat{\tau}$} & \multicolumn{1}{c}{DE} \\ 
  \hline  &&&&\\
    \quad Variable tratamiento\textsuperscript{1} &&&& \\\hline &&&& \\[-1ex]
$pami_i$ &   1 & -0.1*** & 0.03 & -0.1*** & 0.03 \\ 
   &   2 & -0.08** & 0.03 & -0.08** & 0.03 \\ [1.8ex]
    \quad Tipos de seguro\textsuperscript{2} &&&& \\\hline &&&& \\[-1ex]
  $seg_i$ &   1 & 0.46  & 0.48 & 0.43  & 0.77 \\ 
   &   2 & 0.57  & 0.48 & 0.57  & 0.51 \\ [1.8ex]
  $segp_i$ &   1 & 0  & 0.33 & 0  & 0.31 \\ 
   &   2 & 0.34  & 0.47 & 0.38  & 0.56 \\ [1.8ex]
  $segm_i$ &   1 & -0.2  & 0.73 & -0.15  & 0.52 \\ 
   &   2 & 0.28  & 0.42 & 0.3  & 0.47 \\ [1.8ex]
\quad Acceso a bienes y servicios de salud\textsuperscript{2} &&&& \\\hline &&&& \\[-1ex]
  $farm_i$ &   1 & -0.77  & 1.3 & -0.72  & 1.2 \\ 
   &   2 & -1.2  & 2.44 & -1.15  & 2.54 \\ [1.8ex]
  $equi_i$ &   1 & -0.75  & 1.69 & -0.61  & 1.64 \\ 
   &   2 & -1.28  & 2.83 & -1.29  & 3 \\ [1.8ex]
  $smed_i$ &   1 & -16132.99  & 38904.62 & -239.44  & 702.37 \\ 
   &   2 & 5.15  & 17.74 & 5.52  & 16.4 \\ [1.8ex]
  $odon_i$ &   1 & 0.14  & 0.4 & 0.15  & 0.39 \\ 
   &   2 & 0.35  & 0.52 & 0.35  & 0.52 \\ [1.8ex]
\quad \ac{GBS}\textsuperscript{2} &&&& \\\hline &&&& \\[-1ex]
  $lgsalud_i$ &   1 & 21.35  & 232.61 & -2.12  & 7.14 \\ 
   &   2 & 28.32  & 108.63 & 17.13  & 185.52 \\ [1.8ex]
  $pgsalud_i$ &   1 & 0.62  & 29.75 & -0.01  & 4.03 \\ 
   &   2 & -0.46  & 0.57 & -0.46  & 0.54 \\ [1.8ex]
  $cat10_i$ &   1 & -0.92  & 1.09 & -1.3  & 1.98 \\ 
   &   2 & -0.24  & 0.91 & -0.12  & 0.94 \\ [1.8ex]
  $cat25_i$ &   1 & -112.5  & 1812.65 & 3.48  & 29.73 \\ 
   &   2 & -0.23  & 0.73 & -0.24  & 0.8 \\ [1.8ex]
   \hline\hline
\end{tabular}

%% file: tablas/rdd.rkd.bws.tex
\begin{tabular}{lrrrrrrrrr}
  \hline\hline
  && \multicolumn{4}{c}{\ac{RDD}} & \multicolumn{4}{c}{\ac{RKD}} \\\cline{3-10}
  && \multicolumn{2}{c}{Con covariables\textsuperscript{3}} & \multicolumn{2}{c}{Sin covariables} & \multicolumn{2}{c}{Con covariables\textsuperscript{3}} & \multicolumn{2}{c}{Sin covariables} \\\cline{3-10}
Variable objetivo & $p$ & BWL & BWR & BWL & BWR & BWL & BWR & BWL & BWR \\ 
  \hline  &&&&&&&&& \\
      \multicolumn{3}{l}{\quad Variable tratamiento\textsuperscript{1}} &&&&&&& \\\hline &&&&&&&&& \\[-1ex]

$pami_i$ &   1 & 3.80 & 6.72 & 3.80 & 6.66 & 3.15 & 5.33 & 3.16 & 5.29 \\ 
 &   2 & 8.05 & 10.48 & 8.03 & 10.26 & 7.52 & 9.08 & 7.50 & 8.88 \\ [1.8ex]
       \multicolumn{3}{l}{\quad Tipos de seguro\textsuperscript{2}} &&&&&&& \\\hline &&&&&&&&& \\[-1ex]

  $seg_i$ &   1 & 7.24 & 13.23 & 7.66 & 10.85 & 9.61 & 10.13 & 11.04 & 10.02 \\ 
  &   2 & 11.63 & 17.23 & 12.23 & 15.72 & 14.52 & 14.77 & 14.64 & 14.72 \\ [1.8ex]
  $segp_i$ &   1 & 5.80 & 10.87 & 7.73 & 12.09 & 8.67 & 9.67 & 8.47 & 9.95 \\ 
  &   2 & 11.24 & 14.27 & 14.88 & 13.23 & 13.70 & 12.28 & 13.93 & 11.91 \\ [1.8ex]
  $segm_i$ &   1 & 7.63 & 11.52 & 10.44 & 12.82 & 12.06 & 10.58 & 11.06 & 10.29 \\ 
   &   2 & 11.72 & 15.47 & 15.22 & 13.81 & 14.44 & 12.62 & 14.40 & 12.24 \\[1.8ex]
          \multicolumn{7}{l}{\quad Acceso a bienes y servicios de salud\textsuperscript{2}} &&& \\\hline &&&&&&&&& \\[-1ex]

  $farm_i$ &   1 & 7.10 & 10.51 & 7.32 & 11.64 & 8.77 & 8.30 & 8.82 & 8.74 \\ 
   &   2 & 12.92 & 13.21 & 13.03 & 12.37 & 14.98 & 11.44 & 15.00 & 11.25 \\ [1.8ex]
  $equi_i$ &   1 & 8.59 & 13.22 & 8.43 & 12.51 & 11.89 & 10.51 & 11.31 & 10.12 \\ 
   &   2 & 13.26 & 12.88 & 14.47 & 12.98 & 15.30 & 11.26 & 15.49 & 11.33 \\ [1.8ex]
  $smed_i$ &   1 & 9.92 & 10.80 & 10.89 & 10.68 & 13.50 & 9.31 & 13.51 & 9.35 \\ 
 &   2 & 14.25 & 11.70 & 15.14 & 11.57 & 16.69 & 11.21 & 16.69 & 11.32 \\ [1.8ex]
  $odon_i$ &   1 & 9.20 & 12.71 & 7.66 & 12.71 & 6.55 & 10.38 & 6.66 & 10.69 \\ 
   &   2 & 14.19 & 16.61 & 13.90 & 16.44 & 13.29 & 14.60 & 13.33 & 14.62 \\ [1.8ex]
             \multicolumn{3}{l}{\quad \ac{GBS}\textsuperscript{2}} &&&&&&& \\\hline &&&&&&&&& \\[-1ex]

  $lgsalud_i$ &   1 & 7.55 & 8.21 & 7.13 & 10.96 & 12.33 & 8.42 & 9.66 & 8.95 \\ 
   &   2 & 12.78 & 11.40 & 13.04 & 10.62 & 16.87 & 10.16 & 15.87 & 9.35 \\ [1.8ex]
  $pgsalud_i$ &   1 & 7.27 & 7.99 & 9.28 & 8.89 & 10.72 & 7.07 & 10.72 & 6.67 \\ 
   &   2 & 15.39 & 12.52 & 17.10 & 10.21 & 20.05 & 9.16 & 19.70 & 8.92 \\  [1.8ex]
  $cat10_i$ &   1 & 6.92 & 7.72 & 8.76 & 9.61 & 8.93 & 7.84 & 9.42 & 7.40 \\ 
  &   2 & 11.38 & 12.77 & 11.43 & 11.51 & 12.60 & 10.13 & 12.43 & 9.78 \\ [1.8ex]
  $cat25_i$ &   1 & 8.05 & 8.22 & 13.18 & 10.00 & 12.52 & 8.33 & 13.12 & 8.22 \\ 
   &   2 & 13.71 & 15.38 & 16.18 & 15.57 & 16.74 & 13.56 & 16.89 & 13.37 \\ [1.8ex]
   \hline\hline
\end{tabular}